# Current-induced four-state magnetization switching by spin-orbit torques in perpendicular ferromagnetic trilayers

Y. Sheng,[1,2] Y. C. Li,[2,3] X. Q. Ma,[1] K. Y. Wang[2,3,4†]

[1]*Department of Physics, University of Science and Technology Beijing, Beijing 100083, P. R. China*

[2]*State Key Laboratory of Superlattices and Microstructures, Institute of Semiconductors, Chinese Academy of Sciences, Beijing 100083, P. R. China*

[3]*College of Materials Science and Opto-Electronic Technology, University of Chinese Academy of Science, Beijing 100049, P. R. China*

[4]*Center for Excellence in Topological Quantum Computation, University of Chinese Academy of Science, Beijing 100049, P. R. China*

We demonstrated current-induced four-state magnetization switching in a trilayer system using spin-orbit torques. The memory device contains two Co layers with different perpendicular magnetic anisotropy, separated by a space layer of Pt. Making use of the opposite spin current at the top and bottom surface of the middle Pt layer, magnetization of both Co layers can be switched oppositely by the spin-orbit torques with different critical switching currents. By changing the current pulse forms through the device, the four magnetic states memory was demonstrated. Our device provides a new idea for the design of low power and high density spin-orbit torque devices.

[†] Author to whom correspondence should be addressed. Electronic mail: kywang@semi.ac.cn

Spin-orbit torques (SOTs) induced magnetization switching, as an effective way to manipulate spins by electric current, has attracted considerable attentions in recent years due to its high speed and low power advantages.[1–8] Manipulating spins by spin-orbit torques was first observed in ferromagnetic semiconductors[9,10] and subsequently demonstrated in the normal metal (NM)/ferromagnetic metal (FM) multilayers.[3,5–7,11,12] In NM/FM system, NM layer is used as the source of spin current through spin Hall effect, which converts the in-plane current in NM layer to a vertical spin current that exerts torques on adjacent FM layer. Besides spin Hall effect, the Rashba effect, coming from structure inversion asymmetry, changes the spin orientation normal to the current direction.[1,5,8,13] Spin Hall effect was recognized as the main factor determining the switching direction for samples with perpendicular magnetic anisotropy under a fixed magnetic field along the current direction.[14–17] In such NM/FM structure, only the spin current from one side of the NM layer has been exploited to manipulate spins, while the spin current from the other side is wasted.[4,5,18–23] If the spin current at both interfaces can be used to switch the magnetization in two adjacent FM layers, the power consumption of switching two FM layers will be at the similar level with switching one FM layer. However, manipulating magnetic states of two FM layers allows us to realize four-states memory instead of two-state memory, which could be very important for developing high-density memories.[21–25]

In this work, current induced four magnetic states switching has been achieved in Ta (3 nm)/Pt (1)/Co (0.5)/Pt (4)/Co (0.6)/Pt (1.5) structure, where the number in brackets is the layer thickness of nm. Due to large thickness of the spacer Pt (4 nm) layer, almost no ferromagnetic or anti-ferromagnetic coupling was detected between the two ferromagnetic Co layers. Two Co layers, in contact with the opposite interface of the Pt layer, encounter opposite spin-orbit torques when an electric current passing along the film. As a result, the current induced switching behavior of two Co layers are opposite under a fixed magnetic field applied along the current direction. With fixed in-plane magnetic field applied along current orientation, the critical switching current density is proportional to the thickness of ferromagnetic layer and the effective anisotropy field.[26] As both the SOTs and perpendicular magnetic anisotropy are different in two FM layers, the threshold current of magnetization switching of the two FM layers has 8 times difference. Through applying different forms of current pulses to the device, the magnetic state can be switched between four states. By taking advantage of spin current from both Pt interface to switch the two FM layers, we can design four-state memory cell, which can not only increase the storage density but also decrease the energy consumption.

The sample film used in this experiment was Ta (3 nm)/Pt (1)/Co (0.5)/Pt (4)/Co (0.6)/Pt (1.5) grown on a thermally oxidized silicon substrate by magnetron sputtering with thickness of nm in brackets, as shown in Fig. 1a. The bottom Ta serves as a buffer layer which guarantees the perpendicular magnetic anisotropy of the bottom Co layer even the thickness of the Pt layer between them is only 1 nm.[27] Material growth temperature was about 20 °C, with base vacuum better than $10^{-8}$ Torr. Layers of Ta, Co and Pt were deposited by DC sputtering mode and the deposition rates were 0.018 nm/s, 0.012 nm/s and 0.016 nm/s, respectively. During sputtering, the substrate spun at a rate of 10 cycles per minute to ensure uniform deposition of the film on the substrate. Using photolithography and lift-off technique, we fabricated Hall bar devices with conductive channel width of 4 μm. The schematic structure of the Hall bar device is shown in Fig. 1a. The Hall bar structure allows us to investigate the magnetic state of the device using the anomalous Hall effect (AHE).

Fig. 1b shows the change of AHE resistance ($R_{Hall}$) with the vertical magnetic field sweeping from 90 Oe to -90 Oe and then back to 90 Oe, which directly reflects the variation of magnetization normal component. A small current of 0.1 mA was used for detecting the magnetic states while minimized the heating effects. Two steps of AHE resistance change was obtained when the magnetic field was swept from 90 Oe to -90 Oe, and vice versa, where each step is attributed to the magnetization switching from one of the two magnetic layers. Before distinguishing the step of the AHE resistance change corresponding to the layer of the magnetization switching, we identify whether there is interlayer exchange coupling between the two Co layers. The device was firstly magnetized by field of 100 Oe (-100 Oe) along Z axis, which makes sure the magnetization of the two Co layers along the magnetic field direction. Then the magnetic field was swept in a small range of ±16 Oe, which is only large enough to switch the Co layer with smaller switching field. As shown in Fig. 1c, very square one step hysteresis loop was observed for both the 100 Oe and -100 Oe pre-magnetized cases. The switching field of both cases does not change with the pre-magnetized fields, indicating no significant magnetic coupling between these two ferromagnetic layers, which is consistent with the weak interlayer exchange coupling strength of Pt.[28] We then identify which Co layer corresponding to which switching field. A series of reference samples Ta (3 nm)/Pt (1)/Co (0.6)/Pt (4)/Co ($t_{top}$)/Pt (1.5) with $t_{top}$=0.5 nm, 0.6 nm and 0.7 nm were deposited. As the magnetic coupling of two Co layers are weak, we expected that the switching field of the top Co layer will change with its thickness. Fig. 1d shows that the first switching happened at almost the same field of 18 Oe, while the second switching field varied with $t_{top}$, indicating that the top Co layer owns larger switching field. As a result, we can confirm that the switching field and the changed amplitude of $R_{Hall}$ of the top Co is 65 Oe ($H_{ctop}$), 0.19 Ω ($\Delta R_{top}$) and the lower Co is 9 Oe ($H_{cbottom}$), 0.12 Ω ($\Delta R_{bottom}$) for the investigated device in Fig. 1b.

As the spin current induced by spin Hall effect in normal metal decreases rapidly as the film thickness decreasing,[16,29] the thickest middle Pt layer is the main source of spin current. Combining with the large different switching magnetic field of the two Co layers, we expected that the critical current of current-induced magnetization switching for these two layers are also different. The spin current flowing into adjacent Co layer from the middle Pt layer are of opposite spin polarization, inducing opposite current-induced effective fields, as shown in Fig. 2a. Thus the switching behavior of the top and bottom Co layer should to be opposite.

To verify the expectation, the current pulse-induced magnetization switching under different fixed magnetic fields along X axis ($H_x$) were investigated. Current pulses ($I_{pulse}$) of width 100 ms and varied amplitudes were applied to the device. Anomalous Hall resistance ($R_{Hall}$) was measured after each pulse at a low current of 0.1 mA to probe the magnetic states of the Co layers, where all the measurements were performed without any initialization for the magnetic layers. As shown in Fig. 2 b and c, the clockwise current-induced magnetization switching was observed with relative large positive $H_x$ ($\geq$ 1000 Oe), while the current-induced magnetization switching became to anti-clockwise with negative $H_x$ applied. The current-induced magnetization switching could be related with the asymmetry domain wall motion, resulting from the chirality broken of the Néel domain wall which caused by the competition of in-plane magnetic field and Dzyaloshinskii-Moriya interaction[15,30]. With small $H_x$ of 20 Oe or -20 Oe applied, no deterministic switching was observed in both cases. With 1000-1400 Oe applied in current direction, a clockwise current-induced four-state magnetization switching was observed. Compared the changed amplitude of the $R_{Hall}$, we can identify the four magnetic states under current-induced magnetization switching. When the current was swept from -10 mA to 10 mA, the magnetic state of ($Co_{top\uparrow}$, $Co_{bottom\downarrow}$) was formed first and then switched to ($Co_{top↑}$, $Co_{bottom↑}$) at 1.2 mA, where the arrows indicate the magnetization orientation. With the current increased up to 7.5 mA, the magnetic state of ($Co_{top↓}$, $Co_{bottom↑}$) was formed and then kept unchanged with increasing the applied current further. As the current swept back from positive, the magnetic state was firstly changed to ($Co_{top↓}$, $Co_{bottom↓}$) at -1.2 mA and then switched to ($Co_{top↑}$, $Co_{bottom↓}$) with current below -7.5 mA. With increasing the external magnetic field to 1800 Oe, although the four-state current induced magnetization switching was observed, the magnitude of $R_{Hall}$ change between $Co_{bottom↓}$ and $Co_{bottom↑}$ is much smaller. Interestingly, with the external magnetic field at 2200 Oe, only two-state magnetization switching was observed. The two switching amplitude of $R_{Hall}$ is 0.18 Ω, similar to $\Delta R_{top}$ (0.2 Ω) and much larger than $\Delta R_{bottom}$ (0.12 Ω), indicating the current only induced magnetization switching of the top Co layer. The bottom Co layer, with smaller perpendicular magnetic anisotropy, was forced to be aligned along in-plane magnetic field direction and has very small contribution to the $R_{Hall}$, as shown in the insert schematics in Fig. 2 form of the pulse, the width of the pulse in (c) is 100 times the real value.

Finally, we investigated the four memory states controlled by successive current pulses. To obtain clear four states under current pulses, in-plane magnetic field of amplitude 1000 Oe was applied. The current pulses are the combination of 10 mA and 5 mA pulses with the same duration of 100 ms. After each pulse, a 0.1 mA current was injected into the device to detect the magnetization state. As shown in Fig. 3, the four memory states (insert schematics in a and b) were demonstrated with positive and negative 1000 Oe applied in current direction. With -1000 Oe field applied, the magnetic state ($Co_{top↓}$, $Co_{bottom↑}$) was obtained after -10 mA pulse, and the opposite state ($Co_{top↑}$, $Co_{bottom↓}$) was obtained after 10 mA pulse, which is consistent with the results shown in Fig. 2. When two pulses of -10 mA and 5 mA were injected successively into the device, state ($Co_{top\downarrow}$, $Co_{bottom\downarrow}$) was obtained. When two pulses of 10 mA→-5 mA were injected into the device, opposite state ($Co_{top\uparrow}$, $Co_{bottom\uparrow}$) was obtained. In the two pulses cases, the magnetization orientation of both layer were set by +(-)10 mA pulse firstly and then the $Co_{bottom}$ layer with smaller critical switching current was switched by -(+) 5 mA with $Co_{top}$ unchanged. After injecting the same form of pulses in to the device, the magnetization switching under 1000 Oe is opposite to that under -1000 Oe, which is shown in Figure 3b. Through applying different forms of current pulses under a fixed external field, the magnetic states switching between the four states can be well controlled in our device.

In conclusion, we have demonstrated a four-state memory structure consisting of Co/Pt/Co multilayers. Spin current generated at both Co/Pt surfaces are utilized to manipulate the magnetic state in Co layers. With a fixed in-plane magnetic field along the current direction, the current induced opposite spin-orbit torques exert to the top and bottom Co layer, respectively. Thus the top and bottom magnetic layer can be switched reversely. The four memory cell was also demonstrated by successively current pulses with fix external magnetic field applied in the current direction. Our demonstrations could be important for the development of low-power consumption and high density SOTs devices.

**Acknowledgements:**

This work was supported by National Key R&D Program of China No.2017YFA0303400 and 2017YFB0405700. This work was supported also by the NSFC Grant No. 11474272, and 61774144. The Project was sponsored by Chinese Academy of Sciences, grant No. QYZDY-SSW-JSC020, XDPB0603, XDPB0802 and K. C. Wong Education Foundation as well.

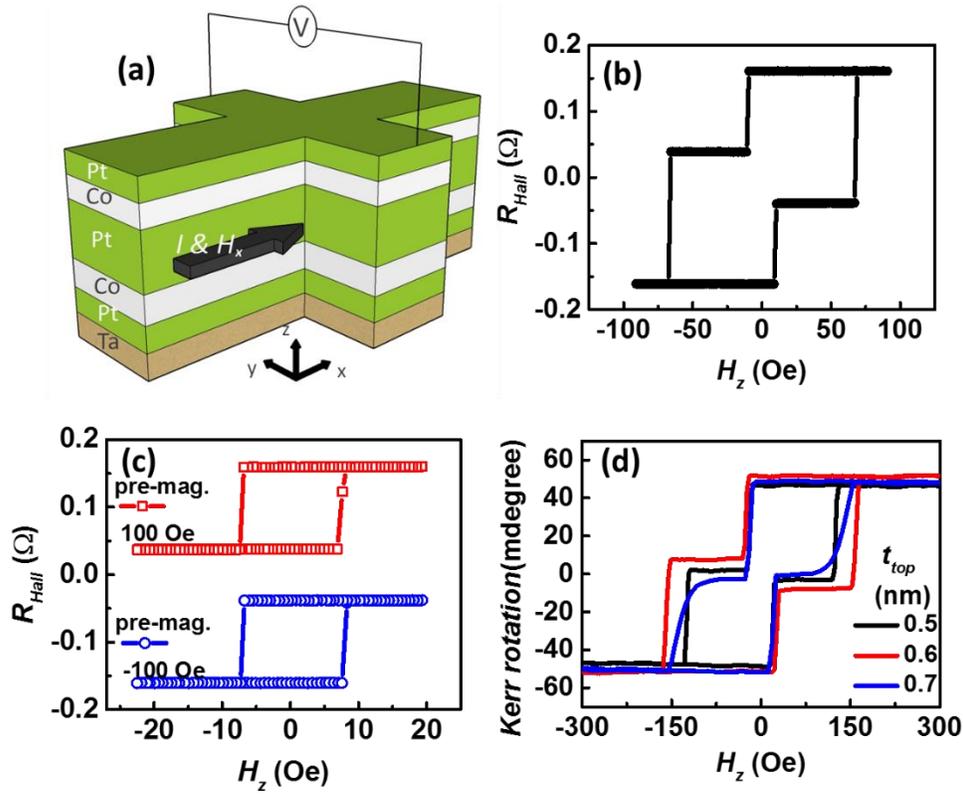

Fig. 1 (a) Schematic of the studied magnetic multilayers system and Hall bar device with the definition of X–Y–Z coordinates. (b) Magnetic hysteresis loops along Z direction measured using the anomalous Hall effect. (c) The minor magnetic hysteresis loops of bottom Co layer after the device was pre-magnetized by 100 Oe (red) and -100 Oe (blue) along Z-direction, respectively. (d) Magnetic hysteresis loops of sample film Ta (3 nm)/Pt (1)/Co (0.6)/Pt (4)/Co ($t_{top}$)/Pt (1.5) measured by polar magnetic-optical Kerr effect, where the thickness of top Co layer $t_{top}$ is 0.5 nm, 0.6 nm and 0.7 nm respectively.

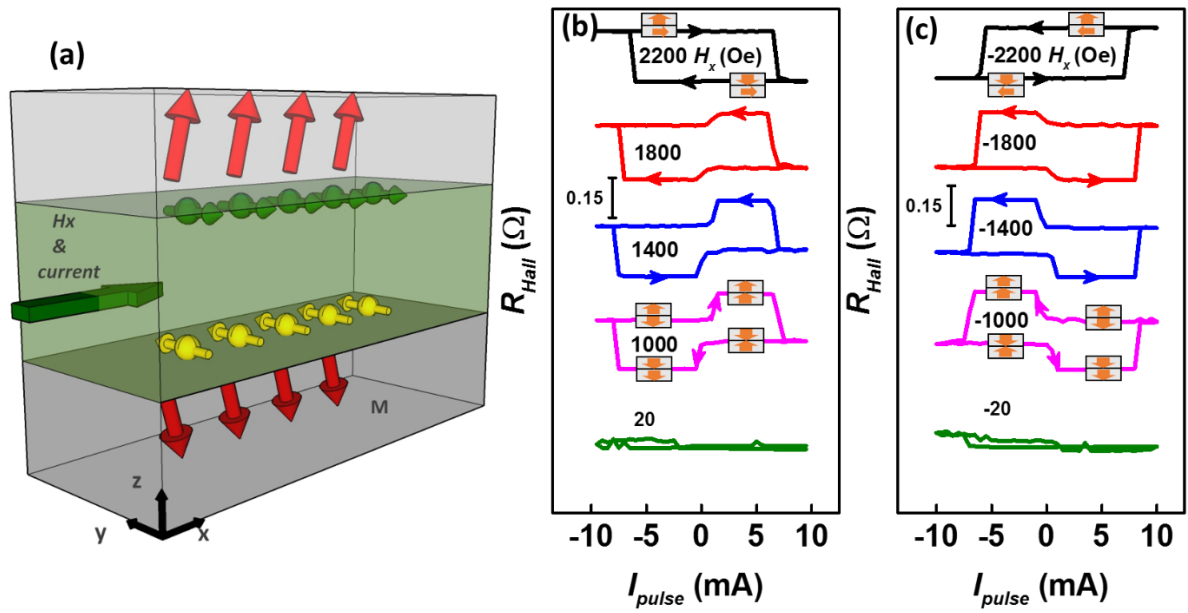

Fig. 2 (a) Illustration of spin Hall effect in the middle Pt layer with magnetic field applied in current direction. (b) and (c) Current pulse induced magnetization switching with positive and negative magnetic field applied along X axis. Schematics inserted in (b) and (c) illustrate the magnetization orientation (orange arrows) of two FM layers at different $R_{Hall}$ levels.

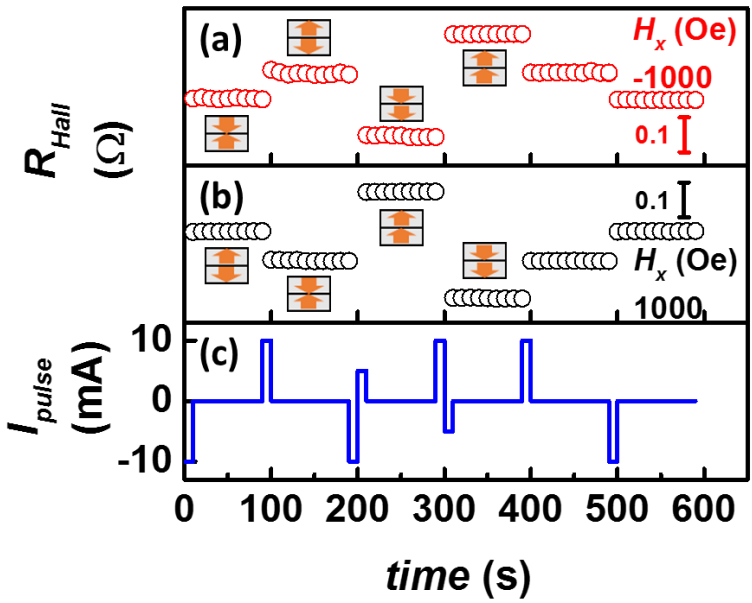

Fig. 3 (a) and (b) Anomalous Hall resistance against time under $H_x$=-1000 Oe and $H_x$=1000 Oe respectively. Schematics inserted in (a) and (b) illustrate the magnetization orientation (orange arrows) of two FM layers at different $R_{Hall}$ levels. (c) Four basic pulses (-10 mA, 10 mA→-5 mA, -10 mA→5 mA, +10 mA) are applied to the device, in which pulses of +-10 mA last for 100 ms, and pulse of (-)+10 mA →-(+)5 mA consists of a pulse of (-)+10 mA and a followed pulse of -(+)5 mA with a total duration of 200 ms. To clearly show the